\documentclass{article}

\usepackage{arxiv}

\usepackage[utf8]{inputenc} 
\usepackage[T1]{fontenc}  
\usepackage{amsmath,amsfonts}
\usepackage{booktabs} 
\usepackage{algorithmic}
\usepackage{algorithm}
\usepackage{xr-hyper} 
\usepackage{hyperref}
\usepackage{array}
\usepackage{nicefrac}       
\usepackage{microtype}      
\usepackage{cleveref} 
\usepackage{textcomp}
\usepackage{stfloats}
\usepackage{caption}
\usepackage{subcaption}
\usepackage{url}
\usepackage{graphicx}
\usepackage{natbib}
\usepackage{doi}

\DeclareUnicodeCharacter{2009}{\,}
\DeclareUnicodeCharacter{2060}{\nolinebreak}
\DeclareUnicodeCharacter{0301}{\'{e}}
\DeclareUnicodeCharacter{03B3}{$\gamma$}

\title{Enhancing Contrast and Resolution for Electron-beam Lithography on Insulating Substrates}

\usepackage{authblk}

\newbox{\orcid}\sbox{\orcid}{\includegraphics[scale=0.06]{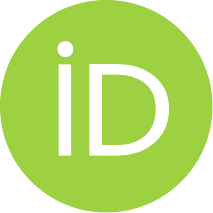}} 
\author[1]{%
	\href{https://orcid.org/0000-0001-9801-8820}{\usebox{\orcid}\hspace{1mm}Deepak Kumar\thanks{\texttt{deepak.kumar@uky.edu}}}%
}
\author[1]{%
	\href{https://orcid.org/0000-0000-0000-0000}{\usebox{\orcid}\hspace{1mm}Cooper Meyers}%
}
\author[1]{%
	\href{https://orcid.org/0000-0000-0000-0000}{\usebox{\orcid}\hspace{1mm}RJ Smith}%
}
\author[1,2]{%
	\href{https://orcid.org/0000-0002-4619-7123}{\usebox{\orcid}\hspace{1mm}J. Todd Hastings\thanks{\texttt{todd.hastings@uky.edu}}}%
}
\affil[1]{Department of Electrical and Computer Engineering, University of Kentucky, Lexington, Kentucky 40506}
\affil[2]{Department of Physics and Astronomy, University of Kentucky, Lexington, Kentucky 40506}

\renewcommand{\shorttitle}{Enhancing Contrast and Resolution for Electron-beam Lithography on Insulating Substrates}

\hypersetup{
            pdftitle={Enhancing Contrast and Resolution for Electron-beam Lithography on Insulating Substrates},
            pdfauthor={Deepak Kumar},
            colorlinks=true,
            linkcolor=blue,
            filecolor=magenta,      
            urlcolor=cyan,
            citecolor=cyan,
}

\begin{document}
\maketitle

\begin{abstract}
We report on the effect of ambient gas on the contrast and the resolution of electron beam lithography (EBL) in gaseous environments on insulating substrates.  Poly(methyl methacrylate) (PMMA) films were exposed in an environmental scanning electron microscope using a 30 keV electron-beam under 1 mbar pressure of helium, water, nitrogen and argon.  We found that the choice of ambient gas results in significant variations in contrast, and the clearing dose increases with the gases’ molecular weight and proton number, consistent with the increase in scattering cross-section. Significantly higher contrast values are obtained for exposure under helium and are accompanied by improved sensitivity. Despite higher sensitivity, helium exhibited the best resolution with 20-nm half-pitch dense lines and spaces.  However, water vapor offered a larger process window, particularly on fused silica substrates.  We also demonstrate that higher sensitivity results from effective charge dissipation. Thus, for EBL on insulating substrates, helium and water vapor may be desirable choices for charge dissipation depending on the substrate and process conditions.
\end{abstract}


\section{Introduction}
Lithography is the process of transferring patterns from one medium to another. Electron beam lithography (e-beam lithography or EBL) uses nanometer-sized focused electron beam for pattern transfer by irradiating substrates that have been coated with an organic or inorganic thin film resist that is sensitive to electrons. The solubility of the resist changes upon exposure to the electrons, allowing for selective removal of either the exposed region (positive-tone resist) or unexposed region (negative-tone resist) in a suitable developer solution. The pattern is finally transferred by either etching or lift-off. 

The primary advantage of EBL over other lithographic techniques is that it provides (i) a maskless patterning process; (ii) high-resolution transfer of complex patterns  with dimensions down to a few nanometers \cite{manfrinato2019patterning}; and (iii) highly automated and precisely controlled operation. Because of its low throughput, EBL is primarily used in integrated-circuit photomask fabrication, small-volume production of semiconductor devices, and research \cite{pala2012encyclopedia}. 

Some EBL applications require electrically insulating substrates including micromechanical systems, unconventional and organic semiconductor electronics, many photonic systems, and nanostructured optical elements.  However, application of EBL for patterning on electrically insulating substrates is often limited because of surface charging effects. Considering the high beam energies employed (typically 30-100 kV) during EBL exposure process, much of the electron dose is deposited in the substrate. When the substrate is electrically insulating, the formation of strong electrostatic fields at the sample surface leads to charged-substrate electron-beam interaction causing the deflection of the incident beam resulting in poor shape fidelity, pattern placement errors \cite{cummings1989charging,arat2019charge}, and even dielectric breakdown of the resist.  Cumming et al. found that the charging problem was dependent on the beam current and independent of the dose, placing an upper limit on the beam current that can be used, hence limiting the write speed \cite{cumming1997efficient}. Recently, pattern-placement error budgets for semiconductor mask writing have become limited by resist charging even on a nominally  conductive EUV masks \cite{shamoun2021multi}. This problem has arisen because charge dissipation layers, such as those discussed below, introduce too many defects and reduce critical dimension control \cite{kim2024improvement}.  

Several methods have been developed to mitigate substrate charging while performing EBL on insulating substrates. Critical energy EBL (CE-EBL) \cite{joo2006nanoscale} takes advantage of the crossover energies between primary electron trapping and secondary electron emission to reduce pattern distortion. However, the resist thickness cannot be more than the penetration depth of the electron beam, and the crossover voltage depends on the substrate and the resist thickness as well as material properties.  Owing to these limitations CB-EBL may not be the most practical option to reduce charging effects.  

The most common method used for charge dissipation in EBL involves coating a thin conductive layer either on top or underneath the electron-beam resist.  Conductive polymer coatings can be effective (see for example references \cite{angelopoulos1993water, huang1994synthesizing,abargues2008charge, dylewicz2010charge}), but can be limited by a shorter shelf life and pH drift.  Conductivity decreases when the pH moves outside the optimal range, increasing the risk of contamination upon precipitation of the polymers \cite{lopez2019charge}.  As noted above, these are being abandonned for current generation EUV mask making.

Thin metal layers  \cite{cumming1997efficient,samantaray2008effect} are also effective for charge dissipation.  Lower atomic number metal layer coatings have been found to significantly alter the clearing dose, but only slightly alter contrast and do not substantially degrade resolution \cite{samantaray2008effect}.  Carbon films deposited by physical and chemical vapor deposition have also been considered for reducing resist charging on mask blanks.\cite{lin2013method}  Film deposition prior to exposure and film removal prior to development can be time consuming, thus impacting the throughput.  In addition, it is important to consider is the method used to deposit the metal layer. When compared to a thermally deposited aluminum layer, electron beam evaporation results in edge roughness that is more than three times higher and also increases the sensitivity of the resist to small dose variations \cite{hambitzer2017comparison}. X-rays and electrons generated in the electron beam evaporator during deposition can also expose the resist \cite{mccord1997electron}.

Finally, variable-pressure electron-beam lithography (VP-EBL), which employs an ambient gas at subatmospheric pressures to reduce charging during EBL on electrically insulating substrates, has been found to be an efficient method for charge dissipation. Previous works demonstrated that low-pressure EBL can eliminate distortion and improve resolution when patterning PMMA on conducting \cite{paul1997effects} and insulating substrates \cite{myers2006variable}; there is a decrease in linewidth variability as chamber pressure increases. When using VP-EBL, there exists a trade-off between resist charging and decreasing linewidth dimensions \cite{paul1997effects}. Recently it was shown that the introduction of water during the EBL exposure process, modifies the chemical processes during e-beam irradiation of Teflon AF \cite{sultan2019altering} and also alters the sensitivity and contrast of PMMA on conductive substrate \cite{kumar2023effect}. However, no data is available on how the ambient gas affects the contrast of the process and the resolution of highly dense patterns on insulating substrates. 

In this work, we studied the effect of ambient gas on the contrast and the resolution of highly dense patterns under various gases when patterning PMMA on insulating substrates. To our knowledge, these are the first studies of molecules other than water for EBL in gaseous environments. Choice of ambient gas results in significant variations in contrast. Clearing dose increases with the gases’ molecular weight and proton number, consistent with the increase in scattering cross-section. High-resolution studies indicated that, despite higher sensitivity, helium exhibited the best resolution with 20-nm half-pitch dense lines and spaces. To our knowledge, this is the highest resolution demonstrated to date for EBL in a gaseous environment.

\section{Experimental details}

\subsection{PMMA spin coating}
PMMA (950 K molecular weight, MicroChem Corp.) was diluted using anisole (MicroChem Corp.) to make 4 wt. \% and 1 wt. \% solution for contrast and resolution experiments respectively. The PMMA solution thus prepared was spin coated onto a n-type $\langle$100$\rangle$ silicon, fused silica and soda lime glass substrate at 500 rpm for 5 s to give a uniform layer and then spun at 4000 rpm for 1 min to set the desired thickness. Next, the spin-coated substrates were heated on a hot plate at 180°C for 120 s to remove any residual solvent. Ellipsometry (M-2000, J. A. Woollam Co. Inc.) was used to measure the final film thickness of the spin-coated PMMA film. The final resist thickness for the contrast and resolution experiments were measured to be 284 nm and 39 nm, respectively.

\subsection{Variable-pressure electron-beam lithography process}
An ELPHY Plus pattern generator (Raith GmbH) coupled with a FEI environmental scanning electron microscope (Quantum FEG 250) with a fast beam blanker was used for the VP-EBL process. A working distance of 10 mm and a beam energy of 30 keV were used for all lithographic processes. A Faraday cup and a pico-ammeter (Keathley 6487) were used for the beam current measurements under vacuum conditions prior to each lithographic exposure.

First, contrast curves were acquired for resist exposure under helium, water, nitrogen, and argon on fused silica and soda lime glass substrates. For contrast experiments, 20 × 100 $\mu$m$^2$ rectangular structures were exposed under 1 mbar pressure for each gas with areal exposure doses ranging from 10 – 300 $\mu$C cm$^{-2}$ with a step size of 12.8 nm and a beam current of 89 pA. An electron-beam energy of 30 keV was chosen to reduce beam scattering in both gas and the resist, as well as to distribute backscattering to the largest range possible \cite{newbury2002x}. Each adjacent rectangle was given a 20 $\mu$m spacing to minimize proximity effects from backscattering. To minimize any variations between exposures, patterns were placed on the same sample 400 $\mu$m apart from each other. Exposed films were developed in methyl isobutyl ketone and isopropyl alcohol (1:3 MiBK:IPA) for 60 s at room temperature followed by 30 s IPA rinse. The thicknesses of the resulting resist after development were measured using a Dektak 6M (Veeco, Inc.) surface profiler.

The second set of experiments was conducted to find the resolution limits for exposure under gaseous environment. For the resolution experiments, “nested-L” structures with 15, 20, 25, 50, 100 and 200 nm half-pitch were exposed under 1 mbar of ambient gases with line dose ranging from 50 – 700 pC cm$^{-1}$ in the increments of 50 pC cm$^{-1}$ with a step size of 3.2 nm and a beam current of 68 pA. Each adjacent "nested-L" pattern was spaced by  4 $\mu$m. The exposed film was developed in 4:1 ethanol:water for 45 s at 3 °C followed by 10 s IPA rinse.  Cold development of PMMA has been demonstrated to provide improved resolution with reduced line edge roughness \cite{ocola2006effect}. After exposure and development, the sample was sputter coated with 5 nm gold. An FEI environmental scanning electron microscope (Quantum FEG 250) and imageJ\cite{schneider2012nih} were used for the line-width measurements.

\section{Results and discussions}

\begin{figure}[htp]
    \centering
    \includegraphics[width=0.75\columnwidth]{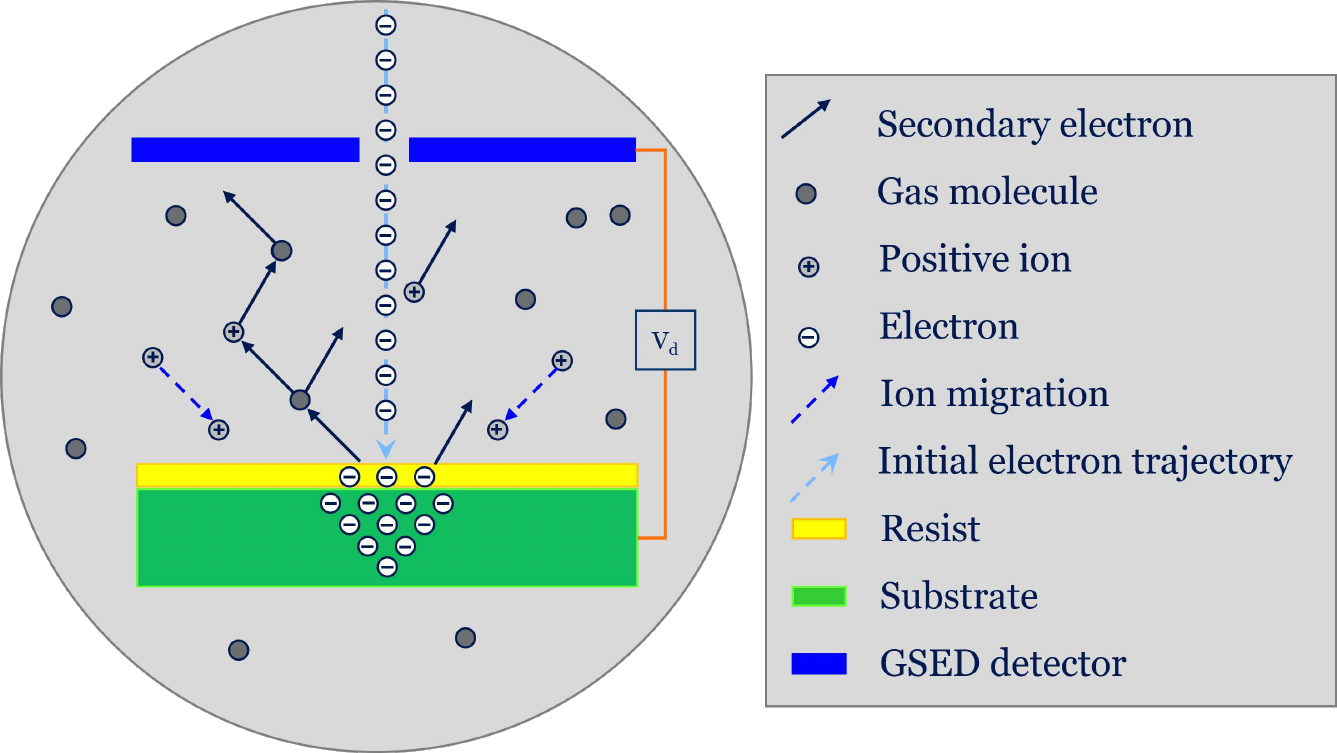}
    \caption{Schematic of VP-EBL. When performing EBL on insulating substrates, the bulk of the sample charges negatively by trapping primary electrons.  Introducing a gas in the chamber can mitigate this charging process because the gas molecules may be ionized by electron impact.  Typically multiple ions are generated through a cascade as secondary electrons are accelerated toward to the biased detector.}
    \label{fig:VP-eBL}
\end{figure}

The use of VP-EBL to pattern under gaseous environment, shown in figure \ref{fig:VP-eBL}, involves the interactions of incident electrons with the gas and the subsequent formation of positive ions, mostly by secondary electrons, which affect the exposure process in several ways: (i) scattering of electrons in the gas could lead to a reduced effective absorbed dose in the resist, and (ii) incident primary electrons slow down under the applied detector voltage, resulting in a decreased landing energy which affects the beam penetration, as well as the interaction volume, electron emission, and charging \cite{stokes2008principles}. Forward scattering in gas significantly affects the exposure process, Danilatos \cite{danilatos1988foundations} showed that the beam skirt radius from plural scattering in a gas is proportional to the effective atomic number of the scatterer; effective atomic number for helium, water, nitrogen, and argon are 2, 7.42, 7 and 18, respectively. For the conditions considered here, electron scattering in the gas results in an absorbed dose, $D_{abs}$, within the primary beam spot of 

\begin{equation}
    D_{abs\_He} > D_{abs\_N_{2}} > D_{abs\_H_2O} > D_{abs\_Ar}
\end{equation}
 
During the VP-EBL exposure process, the secondary electrons emitted from the sample are accelerated toward a positively biased detector and multiplied as they ionize the ambient gas. The secondary electron cascade is collected by the detector and the resulting ions are accelerated toward the grounded or negatively charged substrate.  The multiplication factor $g$, and thus the approximate number of ions per secondary electron generated \cite{thiel2004master}, is given by

\begin{equation}
    g = e^{\alpha d}
    \label{eq:multiplication}
\end{equation}

\noindent where $d$ is the detector-sample distance and $\alpha$ is the first Townsend ionization coefficient which is given by

\begin{equation}
    \alpha = A\cdot P e^{-B\cdot P\cdot d\cdot V_0}
    \label{eq:ionization}
\end{equation}

\noindent where $A$ and $B$ depend on the gas, $V_0$ is the applied detector voltage, and $P$ is the pressure. $P$, $V_0$ and $d$ were kept constant during our experiments; thus, the number of ions per secondary electron generated to balance surface charge is controlled by the choice of gas.  In summary, electron scattering and ion generation in the ambient gas will alter the exposure process; electron scattering in gas reduces the absorbed dose, whilst ion amplification helps to minimize surface charge. 

\subsection{Effect of ambient gas on contrast and sensitivity}

\subsubsection{Sensitivity}
Sensitivity, referred to here as the dose to clear or the clearing dose (D$_C$), is an inherent property of the resist and is defined as the minimum dose at which the exposed resist is soluble in the developer solution. Resist sensitivity depends on a large number of parameters, including resist material, resist thickness, developer, development time, and temperature, as well as the beam energy used for the exposure.  
There is a consistent trade-off between sensitivity and resolution for e-beam resists.  Lower sensitivity is desirable when patterning is limited by shot noise.\cite{ocola2006effect}  Higher sensitivity is desirable to increase throughput and/or minimize the required beam current\cite{mkrtchyan2000global} as well as to reduce temperature rise in the resist\cite{fares2000analytical, yasuda1994resist}.

The sensitivity values of PMMA on soda lime glass and fused silica for exposure under different gases are tabulated in Table \ref{Table: contrast on SLG and FS}. The dose to clear, D$_C$, was obtained by fitting the data in to an empirical model described in detail in reference \cite{ocola2015development}. Soda lime glass differs from fused silica by the fact that it can be conductive at high fields due to the presence of mobile alkali ions \cite{gedeon1999fast, gedeon2000microanalysis}. As previously stated, the sensitivity of resists on insulating substrates for exposure in a gaseous environment is influenced by electron scattering in gas and substrate charging. As is evident from Table \ref{Table: contrast on SLG and FS}, the dose to clear was found to increase with the gases’ molecular weight and proton number, consistent with the increase in scattering cross-section. For the conditions that were investigated in this work, scattering was found to be the dominant contributor influencing the process's sensitivity. However, it is crucial to note that the influence of substrate charging is not inconsequential. When the sensitivity values for the two substrates for helium exposure are compared, the dose to clear is significantly reduced by 15\% when using fused silica instead of soda lime glass.

\begin{table}
\begin{center}
\caption{\label{Table: contrast on SLG and FS}Contrast ($\gamma$) and dose to clear (D$_C$) values of PMMA on soda lime glass and fused silica for exposure under ambient gases.}
\begin{tabular}{| c | c | c | c | c |}
\hline
\multicolumn{1}{|c|}{\textbf{Chamber}}&\multicolumn{2}{c|}{\textbf{Soda lime glass}}&\multicolumn{2}{c|}{\textbf{Fused silica}}\\
\textbf{gas}&D$_C$ ($\mu$C cm$^{-2}$)&$\gamma$&D$_C$ ($\mu$C cm$^{-2}$)&$\gamma$\\ \hline
Helium&130$\pm$0&10.8$\pm$0.4&110$\pm$1&12.4$\pm$0.9\\
Water&148$\pm$0&9.4$\pm$0.3&141$\pm$0&9.9$\pm$0.3\\
Nitrogen&158$\pm$1&8.5$\pm$0.3&145$\pm$0&8.2$\pm$0.2\\
Argon&160$\pm$1&12.3$\pm$0.5&158$\pm$0&8.8$\pm$0.2\\
\hline
\end{tabular}
\end{center}
\end{table} 

\subsubsection{Contrast}

\begin{figure}
    \centering
    \includegraphics[width=0.50\columnwidth]{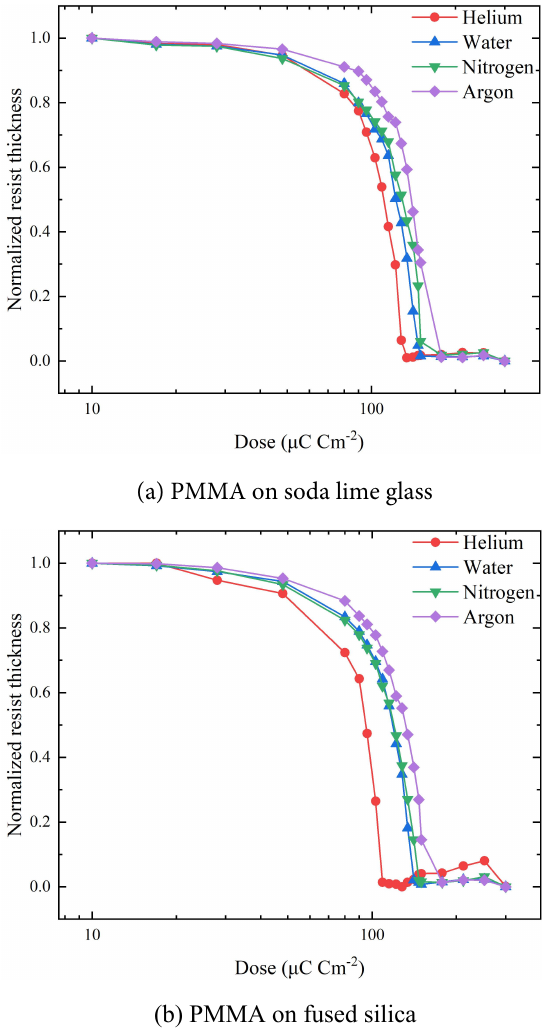}
    \caption{Experimental data showing normalized resist thickness vs exposure dose for resist exposure under ambient gases of PMMA on (a) soda lime glass, and (b) fused silica.}
    \label{Fig: Experimental_contrast_curve}
\end{figure}

The resist contrast, which measures how non-linearly the development process responds to the chemical contrast created in the material after exposure, is crucial in determining the minimum image modulation that can be effectively transformed into a developed resist image \cite{liddle2002resist}. Contrast describes how rapidly developed resist thickness changes with dose, which also has implications upon resolution. Combining contrast and sensitivity, the contrast curve of a resist can be obtained by plotting the residual resist thickness as a function of increasing electron dose. Higher contrast values result in steeper sidewall angles. 

The nature of the development process after the exposure determines the contrast value and is dependent on a number of elements including temperature, development time, developer solvents, and resist properties. Significant enhancement in contrast, at the cost of reduced sensitivity, is obtained for thick PMMA films developed at lower temperatures \cite{rooks2002low}. However, for thin resist films, decreasing the development temperature does not have a discernible impact on contrast while reducing the sensitivity of PMMA to the developer\cite{hu2004sub}. The dose to clear is also greatly influenced by the choice of developer solvents and rinse solution. For PMMA, ethanol/ water in a 4:1 volume ratio offers a non-toxic alternative with superior contrast and resolution over current developers \cite{ocola2015development}. Another method to improve resist contrast uses ultrasonically assisted development; the 3:7 water:IPA composition was found to be optimum for sensitivity and contrast \cite{yasin2001fabrication}. 200 keV EBL exposures lead to higher contrast than 30 keV exposures \cite{duan2010sub}. Contrast gradually increases with increasing water pressure at the expense of reduced sensitivity \cite{kumar2023effect}. 

\begin{figure}
    \centering
    \includegraphics[width=\columnwidth]{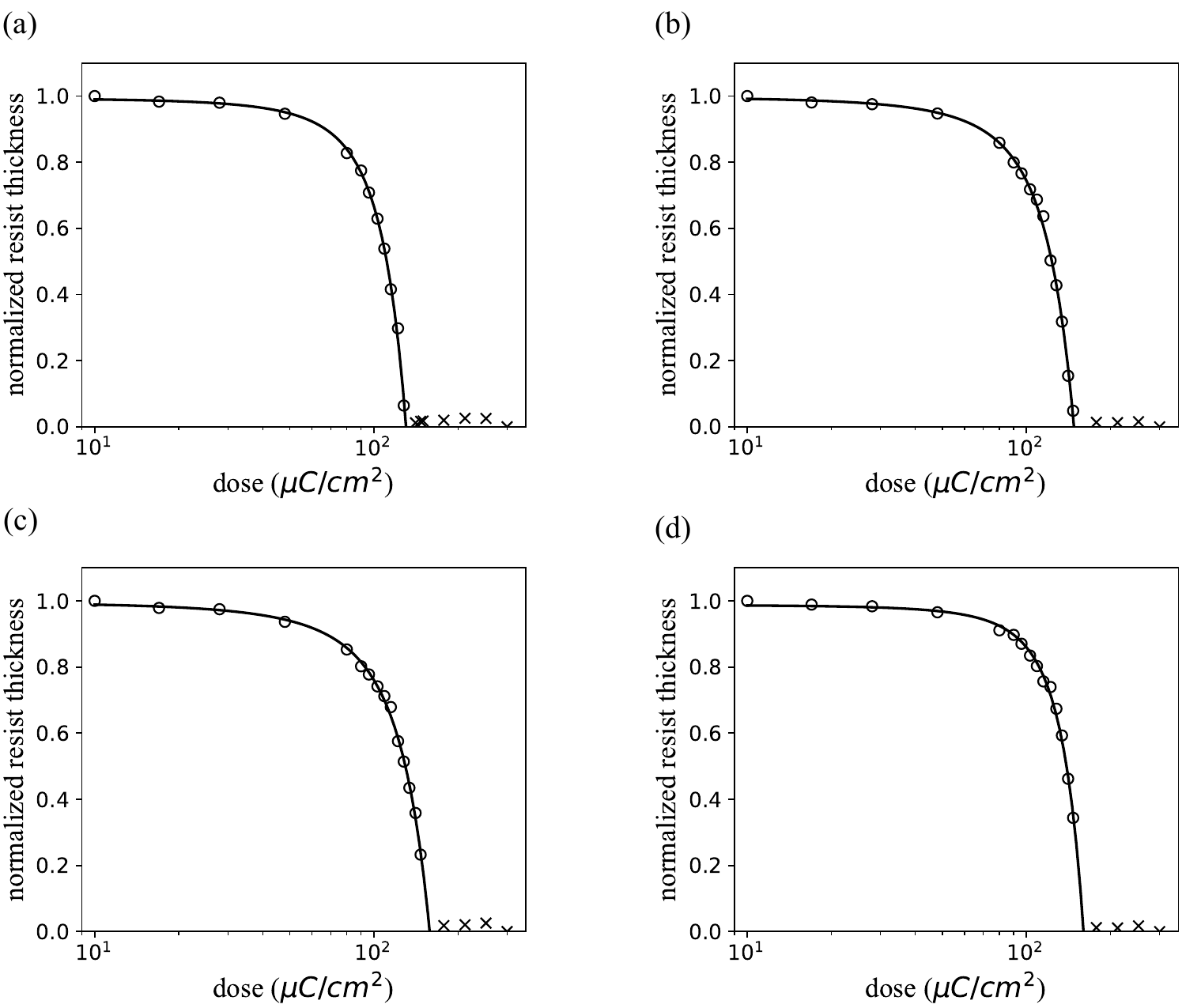}
    \caption{Fitted contrast curve (PMMA on soda lime glass) for exposure under 1 mbar of (a) Helium, (b) Water, (c) Nitrogen, and (d) Argon. Data points used for the fit are indicated by a $\circ$, while data points excluded from the fit are indicated by an ×.}
    \label{fig:fitted contrast curve on soda lime glass}
\end{figure}

\begin{figure}
    \centering
    \includegraphics[width=\columnwidth]{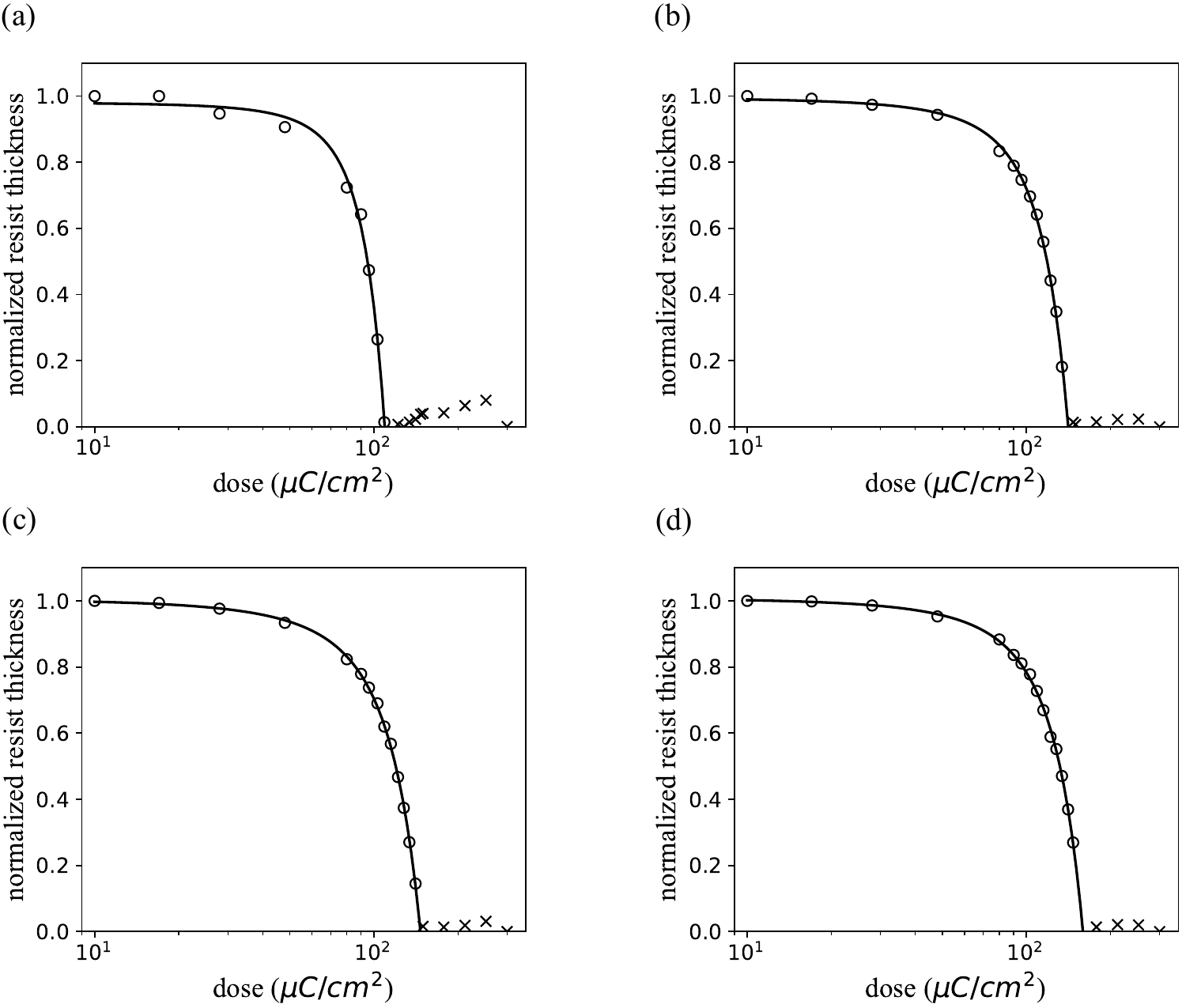}
    \caption{Fitted contrast curve (PMMA on fused silica) for exposure under 1 mbar of (a) Helium, (b) Water, (c) Nitrogen, and (d) Argon. Data points used for the fit are indicated by a $\circ$, while data points excluded from the fit are indicated by an ×.}
    \label{fig:fitted contrast curve on fused silica}
\end{figure}

The contrast curves, normalized residual resist thickness versus areal exposure dose, of PMMA on soda lime glass and fused silica are plotted for different gases in Figure \ref{Fig: Experimental_contrast_curve}. Figures \ref{fig:fitted contrast curve on soda lime glass} and \ref{fig:fitted contrast curve on fused silica} illustrates the fitted contrast curves of PMMA on soda lime glass and fused silica. Data points used for the fit are indicated by a $\circ$, while data points excluded from the fit are indicated by an ×. Table \ref{Table: contrast on SLG and FS} lists the contrast for exposure under different gases. Contrast, $\gamma$, was obtained by fitting the data in to an empirical model described in detail in reference \cite{ocola2015development}. Uncertainties represent the standard error of the fitted parameter. 

Exposing PMMA on insulating substrates in a gaseous environment results in significant variations in contrast. Specifically, resist exposure under nitrogen or argon resulted in improved or degraded contrast values when compared to exposure under helium or water. Higher contrast is obtained for exposure under helium environment, which is also accompanied by substantially reduced dose to clear, which can be correlated to higher absorbed dose and effective charge dissipation. In order to better understand the physical mechanism that results in higher contrast and sensitivity for exposure under helium, we looked into how beam current influences the exposure process. 

\subsection{Effect of beam current on contrast and sensitivity}
We examined the impact of the beam current on contrast and sensitivity to better understand the physical mechanism that leads to increased contrast and sensitivity. The beam current studies were restricted to fused silica substrates in order to remove any complexity related to the high field conductivity on soda lime glass substrates. It is evident from Table \ref{Table: contrast and sensitivity vs beam current on SLG} that the process's sensitivity and contrast are not affected by the beam current when the exposure is carried out under water environment. However, there is a 12\% decrease in the dose to clear at lower beam currents during exposure under helium environment, which can be correlated with effective charge dissipation. Lower beam currents result in higher sensitivity because the substrate has more time to discharge. It is also noteworthy that contrast drastically decreases with the number of exposures conducted under helium environment. This may be due to the resist being exposed by the helium ions. In order to examine the impact of refresh time, which gives the substrate additional time to discharge, we then carried out looped exposures.  

\begin{table}
\begin{center}
\caption{\label{Table: contrast and sensitivity vs beam current on SLG}Contrast ($\gamma$) and dose to clear (D$_C$) of PMMA on fused silica for exposure at different beam currents.}
\begin{tabular}{|c|c|c|c|c|}
\hline
\multicolumn{1}{|c|}{\textbf{Beam}}&\multicolumn{2}{c|}{\textbf{Water}}&\multicolumn{2}{c|}{\textbf{Helium}}\\ 
\textbf{current}&D$_C$ ($\mu$C cm$^{-2}$)&$\gamma$&D$_C$ ($\mu$C cm$^{-2}$)&$\gamma$\\ \hline
78 pA&151$\pm$0&9.9$\pm$0.3&109$\pm$1&6.8$\pm$0.9\\
130 pA&151$\pm$0&10.1$\pm$0.3&112$\pm$1&6.6$\pm$0.4\\
191 pA&150$\pm$0&9.8$\pm$0.2&114$\pm$1&6.6$\pm$0.5\\
292 pA&149$\pm$0&9.6$\pm$0.3&123$\pm$1&7.6$\pm$0.5\\
\hline
\end{tabular}
\end{center}
\end{table}

\subsection{Contrast and dose to clear from loop exposure experiments}

\begin{table}
\begin{center}
\caption{\label{Table: contrast and sensitivity vs beam currents on FS}Contrast ($\gamma$) and dose to clear (D$_C$) of PMMA on soda lime glass and fused silica from loop exposures under water.}
\begin{tabular}{|c|c|c|c|c|}
\hline
\multicolumn{1}{|c|}{\textbf{No. of}}&\multicolumn{2}{c|}{\textbf{Soda lime glass}}&\multicolumn{2}{c|}{\textbf{Fused silica}}\\ 
\textbf{loops}&D$_C$ ($\mu$C cm$^{-2}$)&$\gamma$&D$_C$ ($\mu$C cm$^{-2}$)&$\gamma$\\ \hline
1&140$\pm$0&9.6$\pm$0.2&141$\pm$1&9.0$\pm$0.4\\
2&136$\pm$0&9.8$\pm$0.3&137$\pm$1&9.0$\pm$0.3\\
4&130$\pm$1&9.8$\pm$0.8&135$\pm$0&9.0$\pm$0.3\\
8&123$\pm$0&11.1$\pm$0.3&127$\pm$1&9.0$\pm$0.5\\
\hline
\end{tabular}
\end{center}
\end{table}

The loop exposures were performed under water vapor to eliminate any possible exposure of the resist by helium ions. The fitted dose to clear values vs. the number of loops for PMMA on soda lime glass and fused silica is presented in Table \ref{Table: contrast and sensitivity vs beam currents on FS}. We see that the contrast does not change as the number of loops increases. A 10-12\% reduction in dose to clear for 8 loop exposure is obtained, as would be expected when the substrate has more time to discharge. From these experiments, we conclude that effective charge dissipation tends to reduce the dose to clear. 

\begin{figure}
    \centering
    \includegraphics[width=0.75\columnwidth]{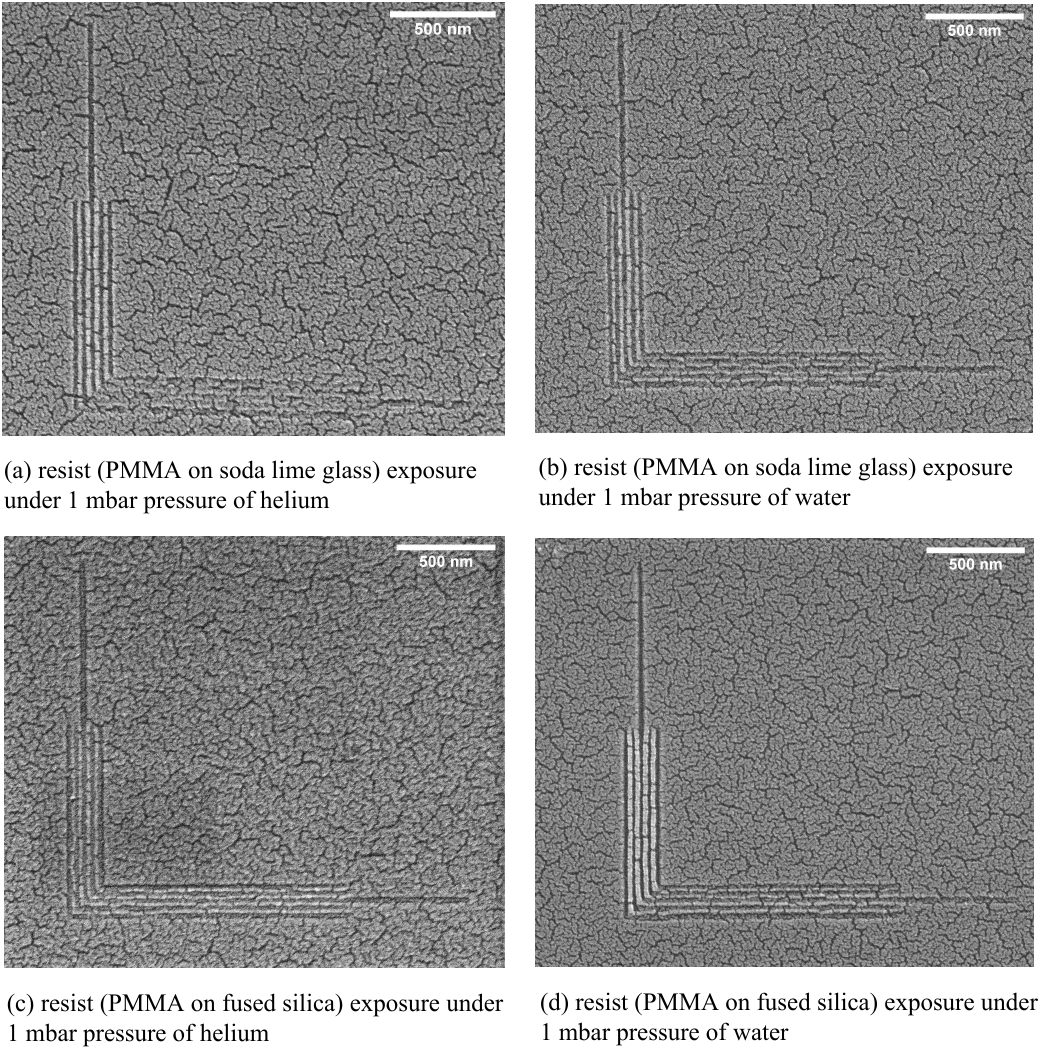}
    \caption{High resolution “nested-L” structures, 20 nm half-pitch; PMMA on soda lime glass exposed under (a) helium and (b) water; PMMA on fused silcia exposed under (c) helium and (d) water.}
    \label{fig:Resolution_SEM_Images_Combined}
\end{figure}

\subsection{High-resolution patterning on insulating substrates}

Resolution defines the minimum feature size or the smallest distance between two patterns that can be resolved. Forward scattering of the beam in the resist and backscattering from the substrate adversely affect the resolution. Resolution is very closely related to both the sensitivity and contrast. It also becomes necessary to consider other factors including exposure charge density, beam accelerating voltage, resist thickness, and resist development process. Fabricating high-resolution sub-10 nm half-pitch features was found to be limited by the resist-development process \cite{duan2010sub}. Resist collapse upon development tends to limit resist film thicknesses to no more than three times the minimum feature size for a typical resist. 

Fig. \ref{fig:Resolution_SEM_Images_Combined} compares high-resolution “nested-L” structures of PMMA on soda lime glass and fused silica for exposure under 1 mbar pressure of water vapor and helium. Exposure of PMMA on soda lime glass under both helium Fig. \ref{fig:Resolution_SEM_Images_Combined} (a) and water vapor Fig. \ref{fig:Resolution_SEM_Images_Combined} (b) exhibited the best resolution with 20-nm half-pitch dense lines and spaces clearly resolved. Exposure of PMMA on fused silica under water, Fig. \ref{fig:Resolution_SEM_Images_Combined} (d), exhibited the best resolution with 20-nm half-pitch. Despite higher sensitivity, exposure under helium (Fig. \ref{fig:Resolution_SEM_Images_Combined} (c)) can still exhibit the best resolution with 20-nm half-pitch at the cost of narrow process window. To the best of our knowledge, these are the highest resolution demonstrated to date for EBL in a gaseous environment.

\subsection{Process window for high-resolution patterning}

\begin{figure}
    \centering
    \includegraphics[width=0.75\columnwidth]{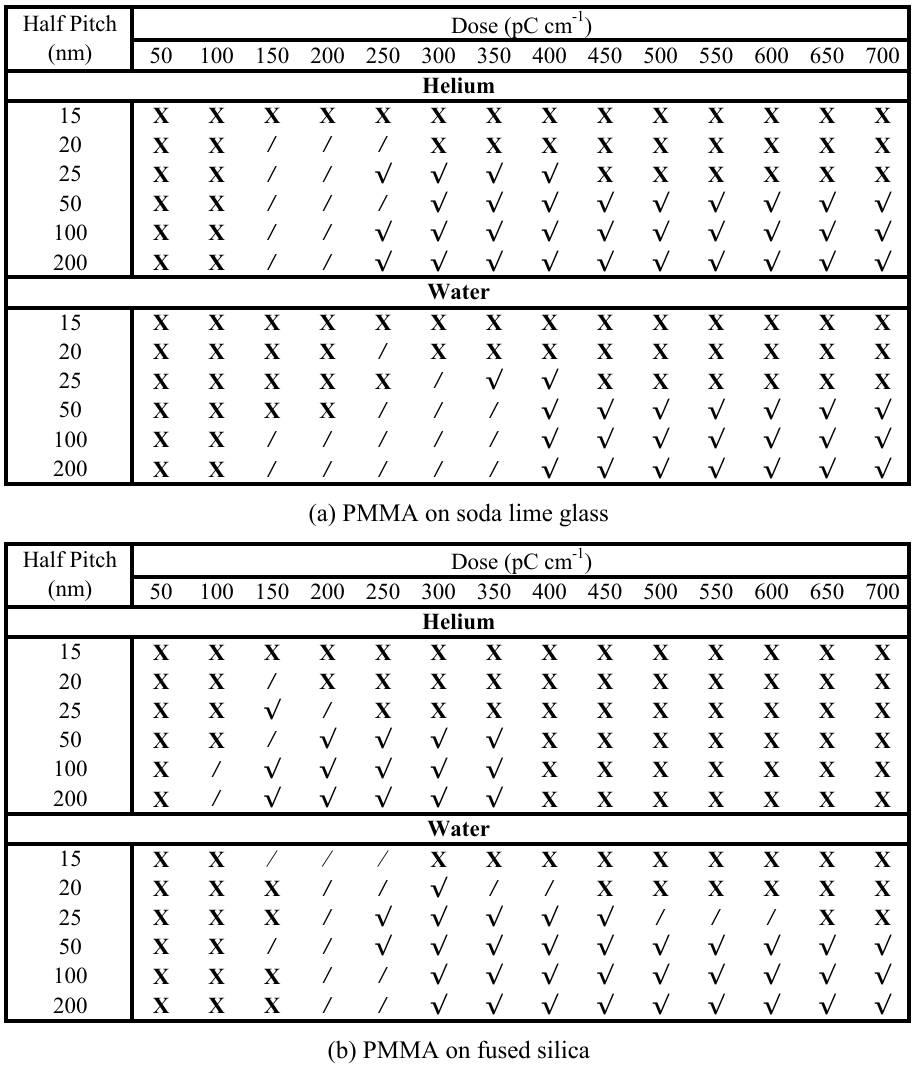}
    \caption{Process window for high-resolution patterning for exposure under helium and water for PMMA on (a) soda lime glass and (b) fused silica. "\textbf{X}" indicates unresolved patterns, "/" indicates underdeveloped or collapsed patterns, and "\checkmark" indicates fully resolved patterns.}
    \label{fig:Resolution_Comparison}
\end{figure}

The process window for high-resolution patterning of PMMA on soda lime glass and fused silica for exposure under 1 mbar pressure of helium and water is compared in Fig. \ref{fig:Resolution_Comparison}. The "\checkmark" symbol here represent the clearly resolvable dense lines and spaces, "/" the underexposed or collapsed patterns, where as, "\textbf{X}" represent the unexposed or overexposed patterns. As can be seen, the process window is wider for exposure under water vapor on fused silica substrates, whereas, the process window is wider for exposure under helium environment on soda lime glass substrates. Thus, for high-resolution patterning under gases, the choice of gas is substrate dependent.

\section{Summary and conclusions}
It was previously established that the introduction of water during electron beam lithography modifies the chemical processes during e-beam patterning of Teflon AF \cite{sultan2019altering} and also alters the sensitivity and contrast of PMMA on conductive substrates\cite{kumar2023effect}. In this work, we studied the effect of ambient gas on the contrast and the resolution of dense patterns for EBL in gaseous environments when patterning PMMA on insulating substrates. To our knowledge, these are the first studies of molecules other than water for EBL in gaseous environments. 

Clearing dose of PMMA was found to increase with the gases’ molecular weight and proton number, consistent with the increase in scattering cross-section. However, variation in contrast values were observed with different gases; especially, exposure under nitrogen and argon results in an improved or degraded contrast depending on the substrate. Regardless of the substrate, significantly higher contrast values are obtained for exposure under helium environment and are accompanied by an improvement in the sensitivity. In addition to the higher absorbed energy in the resist because of the less electron scattering in helium, possible exposure of the resist by helium ions may increase sensitivity. The contrast and dose to clear values for exposure on fused silica are found to improve by around 15\% when comparing exposures on fused silica and soda lime glass in a helium environment.

Experiments studying the impact of beam current on contrast and sensitivity of PMMA on fused silica revealed that there is a 12\% reduction in the dose to clear at lower beam currents during exposure under helium environment; however, contrast drastically decreases with the number of exposures. The improvement in sensitivity and reduction in contrast is correlated to effective charge dissipation and the resist being exposed by helium ions. These are corroborated from results obtained from looped exposures. A reduction of 10-12\% in the dose to clear is obtained with increasing the number of exposure loops. From these results, we conclude that effective charge dissipation tends to reduce the dose to clear. 

High-resolution patterning studies indicated that despite higher sensitivity helium still exhibited the best resolution (20-nm half-pitch lines and spaces) but at the cost of a narrow process window.  This appears to be the highest resolution demonstrated to date for EBL in gaseous environments. The process window for high-resolution patterning on insulating substrates in a gaseous environment is substrate dependent. On fused silica substrates, the process window is wider when exposed under water; on soda lime glass substrates, it is wider when exposed under helium. Thus, on insulating substrates for EBL in gaseous environment, helium yields higher sensitivity without sacrificing resolution.

\section{Acknowledgments}

This work was supported by the National Science Foundation (NSF) under Grant No. CMMI-2135666. This work was performed, in part, at the University of Kentucky Center for Nanoscale Science and Engineering and Electron Microscopy Center, members of the National Nanotechnology Coordinated Infrastructure (NNCI), which was supported by the National Science Foundation (No. NNCI-2025075).

\bibliographystyle{unsrtnat}
\bibliography{References} 

\end{document}